# Microlensing of pulsar radiation in the Galactic Center

N. Wex[1], J. Gil[2], and M. Sendyk[2]

[1] Max-Planck-Society, Research Unit "Theory of Gravitation", University of Jena, Max-Wien-Platz 1, 07743 Jena, Germany
[2] Astronomical Centre, Pedagogical University, Lubuska 2, 65-265 Zielona Góra, Poland

**Abstract.** We investigate the possibility of identifying massive objects (lenses) in the Galactic Center region (GC) by means of pulsar timing. The well known intensity change due to microlensing is found to be less important. For typical stellar masses, the flux magnification can be significant only if the lens passes very close to the pulsar–Earth axis. We show that in the case of a pulsar the time varying travel-time delay, which is observable because of the pulsating nature of pulsar radiation, is a much more powerful tool to investigate mass distributions in the GC. We find that a travel-time delay is measurable even for rather large distances between the lens and the pulsar–Earth axis.

The time varying travel-time delay can be used to determine the mass of the lens and the ratio of the transverse velocity to the minimum impact parameter. Based on Monte Carlo simulations of synthetic data sets we give the expected accuracies in the parameter determination for various masses.

We argue that pulsars found behind the very center of our Galaxy would provide an excellent opportunity to test the mass distribution in the Galactic nucleus, and therefore to distinguish between a super massive black hole (SMBH) and a super-dense star cluster (SDSC) within the central 0.1 pc.

**Key words:** Dark matter – Galactic Center – general relativity – gravitational lensing – pulsars

## 1. Introduction

Star formation rates increase towards the GC (e.g. Biermann 1978). Therefore, one can expect a large number of massive compact objects, such as neutron stars and, perhaps black holes, in central parts of the Galaxy. One should also expect a large number of radio pulsars in this region, although the present known pulsar population does not show this pattern. In fact, when one corrects the known sample of pulsars with respect to various selection effects, then the largest density of pulsars is found near the GC (Lyne et al. 1985, Narayan 1987). There may be as many as $10^5$ pulsars in the vicinity of the actual GC (Lyne et al. 1985). The most important selection effects which bias the sample of discovered pulsars towards nearby sources include background radiation, dispersion smearing, and scattering pulse broadening. These effects can be largely reduced by choosing high observing radio frequencies. Although energy spectra of radio pulsars are dominated by the low frequency regime, pulsars also seem to be quite luminous at high frequencies. Recent detection of a number of pulsars at 35 GHz at the Effelsberg Observatory indicates the importance of the high frequency regime (Wielebinski et al. 1993). It seems that a receiver frequency of about 5 GHz should be a reasonable compromise between the lower flux densities and the largely reduced background radiation, dispersion smearing and scatter broadening. This allows a large receiver bandwidth of about 500 MHz. Detailed considerations for a 100 meter radiotelescope equipped with a low-noise receiver (Fig. 1) show that such a high frequency search would be sensitive to luminous pulsars near the Galactic Center with periods longer than about 100 ms.

Three searches have been undertaken so far at frequencies near 1.5 GHz. A total of 147 new pulsars have been discovered, many of them young and with relatively short periods (Johnston 1990, Clifton & Lyne 1986). The 5 GHz search can avoid scattering broadening for Dispersion Measures (DM) in excess of 1000 cm$^{-3}$pc. Since the base sensitivity for long period signals at 5 GHz is about 0.2 mJy (Fig. 1), as compared with $\sim 1$ mJy in previous searches at 1.5 GHz (Clifton & Lyne 1986, Johnston 1990), one can expect to find new pulsars with relatively flat spectra $S \propto \nu^{-a}$. Approximately 50% of pulsars have such a flat spectra with value of $a$ less than 1.7 and the median value of $a$ for pulsars detected by Johnston (1990) at 1.5 GHz is 1.0. Another high frequency search for GC pulsars at 2.6 GHz is planned for summer 1995 at Parkes radio-observatory (Lyne 1995).

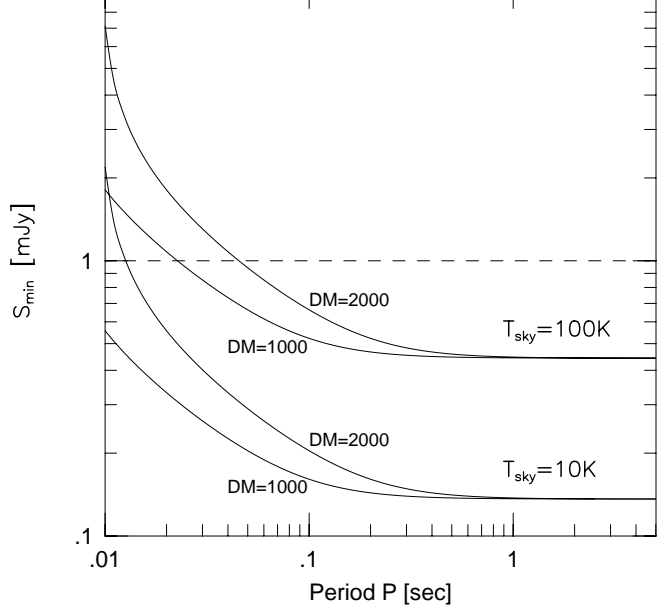

**Fig. 1.** Flux minimum sensitivity curves for a 5 GHz receiver with the system temperature $T_{sys} = 30$ K installed at a 100 m radiotelescope. The curves are calculated for DM exceeding 1000 cm$^{-3}$pc for both cold ($T = 10$ K) and hot ($T = 100$ K) sky. Total receiver bandwidth assumed is 500 MHz. The horizontal line at 1 mJy corresponds to the sensitivity obtained in previous high frequency pulsar surveys.

Small dispersion smearing and negligible scattering at 5 GHz proves to be crucial in probing deeper into the GC. The 5 GHz pulsar search, if conducted with one of the world largest radiotelescopes, has the potential of discovering exotic objects such as newly born pulsars, relativistic pulsar neutron-star or pulsar black-hole binary systems. The number of such systems in the GC may be large. By accurate measurements of the arrival times of the pulses one can determine the mass of the pulsar and of its companion, i.e. if two post-Keplerian parameters are measured the masses of pulsar and companion are known up to the Doppler-factor which is negligible in most cases. This may lead to the discovery of the first black hole (see Taylor (1993), Taylor & Weisberg (1989) for more information on the timing of binary pulsars). Below we discuss the possibility of discovering massive objects in the GC even if they do not occur in binaries with radio pulsars.

When the star radiation passes near a massive object, its intensity is magnified due to the microlensing phenomenon. Several such microlensing effects have recently been observed in the direction of the Baade Window in the Galactic Bulge (Udalski et al. 1994 1994, Kiraga & Paczynski 1994). Similar effect should also apply to pulsar signals. However, because of the pulsating nature of pulsar radiation, the flux magnification will be accompanied by a time kova & Doroshenko 1994). In this paper we come to the conclusion that regularly timed pulsars in the GC can lead to the detection of a number of microlensing events. In the case of lenses of more than a solar mass this can lead to a very precise mass determination.

The motion of the stars and gas clouds surrounding the very center of our Galaxy is most easily explained by a central mass of about $10^6 M_\odot$ within a radius of 0.1 pc (Genzel et al. 1994). The observations of a compact, nonthermal, radio source termed Sgr A$^*$ gives a circumstantial evidence that our Galactic nucleus houses a SMBH. Only recently Narayan et al. (1995) published a model for Sgr A$^*$ in which a black hole of mass $7 \times 10^5 M_\odot$ accretes matter with a rate of $(1.2 \times 10^5)\alpha M_\odot$/yr, where $\alpha$ is a parameter which describes the viscosity of the gas. The model fits the observed spectrum of Sgr A$^*$ from radio to hard X-ray wavelengths. The observation of a pulsar located just behind the Galactic nucleus would provide a unique possibility to identify a SMBH and determine its mass with high accuracy. Thus the discovery of pulsars in the GC could resolve the SMBH-SDSC controversy concerning the very center of our Galaxy.

The paper is organized as following: In Sect. 2 we summarize results concerning the flux-magnification and angular separation caused by a Schwarzschild lens. In Sect. 3 we calculate the time varying travel-time delay caused by a passing Schwarzschild lens. We distinguish between a weak and a strong travel-time delay event. In Sect. 4 we investigate the change of the observed spin parameters of the pulsar (period, period derivative, ...) due to a time-varying travel-time delay. In Sect. 5 we present expected accuracies of the parameter determination in the case of weak travel-time delay. In Sect. 6 we present expected accuracies in the parameter determination in case of strong travel-time delay. In Sect. 7 we compare the travel-time delay caused by a SMBH and the travel-time delay caused by a SDSC in the Galactic nucleus. In Sect. 8 we present statistical investigations. We calculate the probability of observing a time-delay effect toward the GC and the possibility of finding a pulsar behind a possible SMBH in the Galactic nucleus. Finally, we present our conclusions in Sect. 9.

## 2. Flux magnification and angular separation

Let us consider the influence of a compact astronomical object with a mass $M$ on the radiation of a pulsar, when it passes close to the pulsar–Earth axis. This could be an ordinary star, a white dwarf, a neutron star or a black hole and, generally, we shall call it the lens. The corresponding lensing geometry for such a Schwarzschild lens is presented in Fig. 2. PSR denotes the position of the pulsar and SSB denotes the position of the barycentre of the solar system, by what we mean the proper reference frame attached to the mass centre of the Solar System.

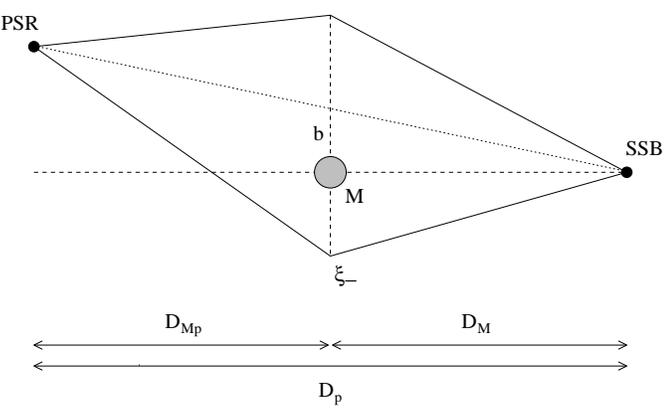

**Fig. 2.** Lensing geometry of the PSR–M–SSB system. Here $b$ is the impact parameter and $\xi_+, \xi_-$ are the distances in the lens plain between the mass and the two different light rays connecting PSR and SSB.

As for ordinary stars the pulsar radiation will experience a flux magnification. If the observed intensity in case of no lens is equal to $I_0$ then the observed intensities of the two images, $I_+$ and $I_-$, obey the following relations: (Refsdal 1964)

$$\mu_+ \equiv \frac{I_+}{I_0} = \frac{1}{4}\left[\frac{f}{\sqrt{f^2+4}} + \frac{\sqrt{f^2+4}}{f} + 2\right],$$

$$\mu_- \equiv \frac{I_-}{I_0} = \frac{1}{4}\left[\frac{f}{\sqrt{f^2+4}} + \frac{\sqrt{f^2+4}}{f} - 2\right], \quad (1)$$

$$A \equiv \mu_+ + \mu_- = \frac{f^2+2}{f\sqrt{f^2+4}},$$

where $f$ is a dimensionless impact parameter

$$f = b/R_E \quad (2)$$

and

$$R_E \equiv \sqrt{2R_S \frac{D_M D_{Mp}}{D_p}} \quad (3)$$

is the *Einstein radius*. Here $R_S$ is the *Schwarzschild radius* of the lens

$$R_S \equiv \frac{2GM}{c^2}. \quad (4)$$

The ratio of the observed intensities of the individual images is

$$r = \frac{I_+}{I_-} = \left(\frac{F_+}{F_-}\right)^2 \xrightarrow{f \gg 1} f^4, \quad (5)$$

where the definition

$$F_\pm \equiv \sqrt{f^2+4} \pm f. \quad (6)$$

If the impact parameter $b$ is much larger than the Einstein radius $R_E$, then the contribution of the second image to the total brightness is completely negligible. For instance if $f > 2.85$ then the contribution of the second image is less than 1%.

In the case when both pulsar and lens are located in the GC region we find $D_{Mp} \ll D_M \approx D_p$. Therefore the Einstein radius is approximated by

$$R_E \approx (2.85 \text{ AU})\left(\frac{M}{M_\odot}\right)^{1/2}\left(\frac{D_{Mp}}{1 \text{ kpc}}\right)^{1/2}. \quad (7)$$

For normal stars acting as lenses in the GC the Einstein radius is a few AU. For a SMBH of $\sim 10^6 M_\odot$ we find an Einstein radius of the order of $10^3$ AU.

The angular separation of the two images is (see Refsdal 1964)

$$\Delta\theta = \frac{R_E}{D_M}\sqrt{f^2+4}. \quad (8)$$

If pulsar and lens are located in the GC we have $D_M \approx 8$ kpc. Using Eq. (7) we find

$$\Delta\theta \approx 3\rlap{.}''6 \times 10^{-4}\sqrt{f^2+4}\left(\frac{M}{M_\odot}\right)^{1/2}\left(\frac{D_{Mp}}{1 \text{ kpc}}\right)^{1/2}. \quad (9)$$

For a SMBH ($M \sim 10^6 M_\odot$) we expect $\Delta\theta$ to be of the order of arc seconds.

## 3. travel-time delay

The pulsating nature of pulsar radiation offers the possibility of observing another aspect of gravitational lensing, namely the change in the light travel-time. There are two effects that contribute to the light-travel time of the pulsar signals (see Cooke & Kantowski 1975, Schneider et al. 1993). First, the deflection of the light ray increases the length of the light path (*geometrical time delay*). Secondly, the light traverses the gravitational field of the lens and therefore suffers a *potential time delay*. If $b \ll D_{Mp}, D_M$, the light-travel time for a Schwarzschild lens is given by (see Appendix A)

$$\tau_\pm = \tau_M\left(4F_\pm^{-2} - 2\ln F_\pm\right) + const. \quad , \quad (10)$$

where we introduced the "Schwarzschild time"

$$\tau_M \equiv R_S/c. \quad (11)$$

Since the lens is moving, the light-travel time depends on the time of observation in a characteristic way. Writing

$$b^2 = b_m^2(1+s^2) \quad (12)$$

we find that

$$f = f_m\sqrt{1+s^2}, \qquad s = q(t-T_0), \quad (13)$$

$$f_m \equiv b_m/R_E, \qquad q \equiv v_\perp/b_m. \qquad (14)$$

Here $T_0$ is the time of the closest approach of the lens to the PSR–SSB line.

In the case of a *weak lensing event* ($f_m \gg 1$) we can neglect the geometrical time delay in Eq.(10) for the first light ray (+). The time delay is then dominated by the potential time delay

$$\tau_+ = -\tau_M \ln(1 + s^2) + const. \qquad (15)$$

In the solar system this is known as the Shapiro delay (Shapiro 1964) which is tested with a measurement precision of 0.1% (see e.g. Will 1992). According to Eq. (1) the intensity of the second image (−) is negligible.

In order to measure the travel-time delays, we need a number of precise measurements of topocentric pulse arrival times (TOAs). Let the parameter $s = s_0$ corresponds to the beginning of TOA monitoring at the initial time $t = t_0$. We define

$$\Delta\tau \equiv \tau(f) - \tau(f_0). \qquad (16)$$

For the brighter image (+) the maximum delay occurs at $s = 0$, corresponding to the closest approach of the lens to the line-of-sight when $b = b_m$

$$\Delta\tau_{\max} = \Delta\tau(f_m). \qquad (17)$$

The time between the beginning of regular observations ($t_0$) and the maximum ($T_0$) written in more suitable units is

$$T_0 - t_0 = -(4.7 \text{ yr}) \, s_0 \, \frac{b_m[AU]}{v_\perp[\text{km/s}]} \equiv -(4.7 \text{ yr}) \, s_0 \, \mathcal{Q}^{-1}. \qquad (18)$$

Let us consider lensing objects with stellar masses $M = 1 M_\odot \ldots 20 M_\odot$. For $b_m \approx 100$ AU, $v_\perp \approx 300$ km/s, and $s_0 = -1$ we find that

$$T_0 - t_0 \approx 1.6 \text{ yr}, \qquad (19)$$

and the corresponding maximum differential Shapiro delay is

$$\Delta\tau_{\max} \approx 7 \ldots 136 \ \mu\text{s}. \qquad (20)$$

A more detailed investigation of Eq. (16) for the case of a weak travel-time delay (cf. Eq. (15)) is presented in Fig. 3.

The actual GC (Sagittarius A West) is likely to contain a SMBH with a mass $M \sim 10^6 M_\odot$. Radio-continuum VLBI observations show a core source not larger than several AU in size. One can therefore expect that radiation of the innermost GC pulsars will be gravitationally affected by the putative SMBH. The Shapiro delay will be strongest for pulsars with small angular separation from the Sag A West. The fact that in this case the lens is at rest and the pulsar is moving does not change the considerations given above. As can be seen from Fig. 3 the Shapiro delay for $M/M_\odot \sim 10^6$ is of the order of several seconds.

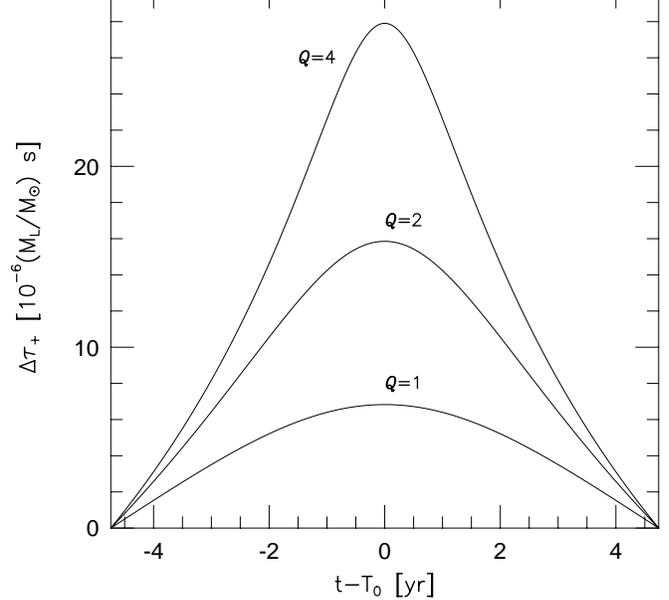

**Fig. 3.** Differential Shapiro delay [in $10^{-6}(M/M_\odot)$ s] as a function of time $t - T_0$ [in years].

## 4. travel-time delay induced period variations

In this section we investigate the influence of the travel-time delay on the observed period $P$ and the observed period change $\dot{P}$ of the pulsar. Explicit equations and figures we shall give only for the weak travel-time delay, see Eq. (15). The influence of a strong travel-time delay is qualitatively the same.

Let us denote the observed period and the observed period change in the case of no lens ("intrinsic") by $P_i$ and $\dot{P}_i$, respectively. To simplify the notation we shall use the definition

$$\mathcal{M} \equiv M/M_\odot. \qquad (21)$$

The difference in arrival time of two consecutive pulses is the observed period of the pulsar, which is given by

$$P \equiv t_2 - t_1 = P_i + \tau(t_2) - \tau(t_1) \simeq P_i + P_i \frac{d\tau}{dt}(t_1). \qquad (22)$$

Since $ds/dt = q$ (Eq. 14) we find

$$P_+ \simeq P_i \left(1 - \tau_M q \frac{2s}{1+s^2}\right). \qquad (23)$$

One can see that $P$ is equal to $P_i$ for $s = 0$ and $s = \pm\infty$ and the deviation of $P$ from $P_i$ is maximum for $s = \pm 1$. Thus

$$\max \left| \frac{P_+}{P_i} - 1 \right| = \tau_M q \simeq 6.58 \times 10^{-14} \mathcal{M} \mathcal{Q}. \qquad (24)$$

For a weak lensing event the change of the period $P_+$ with time is shown in Fig. 4.

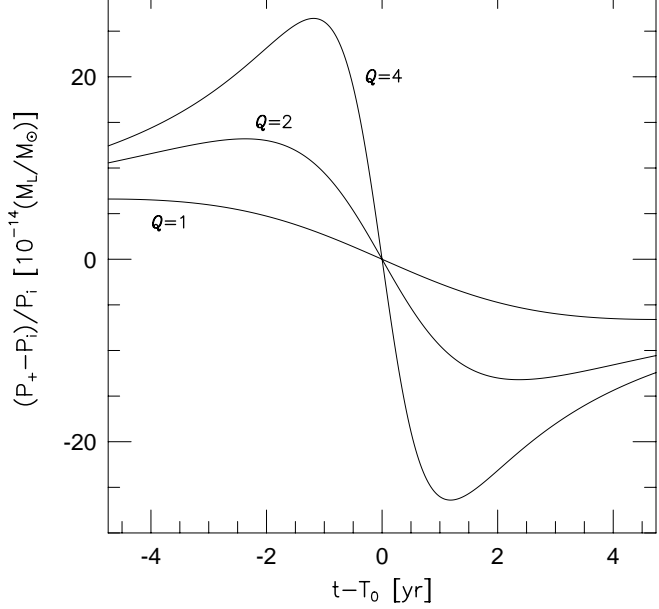 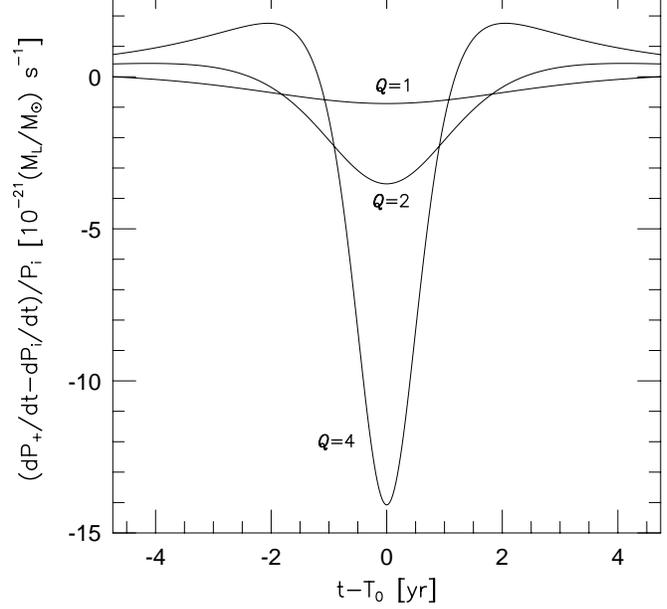

**Fig. 4.** Weak lensing event: $(P_+ - P_i)/P_i$ [in $10^{-14}(M/M_\odot)$] as a function of time $t - T_0$ [in years].

**Fig. 5.** Weak lensing event: $(\dot{P}_+ - \dot{P}_i)/P_i$ [in $10^{-21}(M/M_\odot)$ $\mathrm{s}^{-1}$] as a function of time $t - T_0$ [in years]

The time derivative of the observed period is

$$\dot{P}_+ \simeq \dot{P}_i - P_i \tau_M q^2 \frac{2(1-s^2)}{(1+s^2)^2}. \tag{25}$$

The change of the period due to the travel-time delay is maximum for $s = 0$:

$$\max \left| \frac{\dot{P}_+}{P_+} - \frac{\dot{P}_i}{P_i} \right| \simeq (8.8 \times 10^{-22} \text{ s}^{-1}) \mathcal{M} \mathcal{Q}^2. \tag{26}$$

For example, if $v_\perp = 300$ km/s, $b_m = 100$ AU, and $M = 1 \ldots 20\, M_\odot$ we find that

$$\max \left| \frac{\dot{P}_+}{P_+} - \frac{\dot{P}_i}{P_i} \right| \simeq 7.9 \times 10^{-21} \ldots \; 1.6 \times 10^{-19} \text{ s}^{-1}. \tag{27}$$

Variations of $\dot{P}_+$ with time for a weak lensing event are presented in Fig. 5.

The measurement precision for $\dot{P}/P$ for a typical millisecond pulsar observed for several years is in the range of $10^{-21} \ldots \; 10^{-18}$ s$^{-1}$. Therefore a lensing event in our Galaxy with $b_m \sim 100$ AU might just be observable in the case of a stellar mass.

The situation is completely different for a central SMBH with $M \sim 10^6 M_\odot$, since the mass is scaled by factor of $10^6$. In this case the modulation of pulsar period by microlensing is comparable with a slow-down rate due to the magnetic dipole radiation, which is typically $10^{-15}$ ss$^{-1}$.

Finally, we would like to mention that the time varying travel-time delay can influence significantly the observed second period derivative $\ddot{P}_+$. If $f_m \gg 1$ than $\ddot{P}_+$ is maximum for $s = \pm(\sqrt{2} - 1)$ and

$$\max \left| \frac{\ddot{P}_+}{P_+} - \frac{\ddot{P}_i}{P_i} \right| \simeq (8.6 \times 10^{-30} \text{ s}^{-2}) \mathcal{M} \mathcal{Q}^3. \tag{28}$$

Thus, even a rather weak travel-time delay ($b_m \sim 100$ AU, $v_\perp \sim 100$ km/s) could cause a "braking index"

$$n_B = 2 - \frac{P \ddot{P}}{\dot{P}^2} \tag{29}$$

that is much bigger than 3 ($n_B = 3$ is expected for rotational energy loss by dipole radiation). For example, in case of PSR B1937+21 the observed braking index is about 4500. This can be explained by a travel-time delay caused by a lens of a few solar masses (see Wex 1995). A number of pulsars show the value of $n_B$ considerably larger than 3. Perhaps this is a result of the travel-time delay caused by a massive object passing near the line-of-sight.

## 5. Parameter analysis for a weak lensing event ($f_m \gg 1$)

For lenses with stellar masses ($M = 1 M_\odot \ldots 20 M_\odot$) the Einstein radius is expected to be of the order of a few AU, as shown in Sect. 2. Thus in this case a strong lensing of the pulsar radiation ($f \lesssim 1$) is very unlikely (cf. Sect. 8). Therefore, we will now concentrate solely on weak lensing events to the next section.

In this section we investigate the expected accuracies for the parameter determination in the case of a weak lensing event ($f_m \gg 1$). Since the second image ($-$) is much fainter than the first one ($+$) there is only one pulsar signal for timing, which practically does not change in intensity, i.e. the change in intensity is too weak to allow for some parameter determination. Since the arrival time of pulsar signals can be measured with very high accuracy we have a chance to obtain information about the lens by observing the travel-time delay caused by the lensing mass. We present a *timing-formula* for pulsars that show a weak travel-time delay in their TOAs.

Let us denote the epoch of observation by $t_\oplus$. A weak travel-time delay is characterized by three parameters, $M$, $q$, and $T_0$ (see Eq. 15). We denote the "intrinsic" frequency and its derivatives with respect to time corresponding to the barycentric arrival time $t = t_\oplus$ by $\nu$, $\dot\nu$, $\ddot\nu$, ... (Note: "intrinsic" denotes the frequency and its derivatives measured in the absence of any lensing mass.)

The "intrinsic" frequency corresponding to a certain barycentric arrival time $t$ is

$$\nu_i = \nu + \dot\nu(t - t_\oplus) + \tfrac{1}{2}\ddot\nu(t - t_\oplus)^2 + \ldots \qquad (30)$$

The observed frequency at the time $t$ is

$$\bar\nu = \nu_i \left(1 + \frac{d\tau_+}{dt}\right)^{-1}. \qquad (31)$$

Since $d\tau/dt \sim 10^{-14} \mathcal{MQ}$ we find (neglecting $\ddot\nu$ and higher derivatives of $\nu$)

$$\bar\nu(t) \simeq \nu_i \left(1 - \frac{d\tau_+}{dt}\right) \simeq \nu + \dot\nu(t - t_\oplus) - \nu \frac{d\tau_+}{dt}. \qquad (32)$$

The pulsar phase at the barycentric arrival time $t$ is given by

$$\phi = \int_{t_\oplus}^{t} \bar\nu(t)\, dt \equiv \Phi(t; \phi_0, \nu, \dot\nu, M, q, T_0). \qquad (33)$$

Thus

$$\begin{aligned}\Phi(t; \phi_0, \nu, \dot\nu, M, q, T_0) = \\ \phi_0 + \nu(t - t_\oplus) + \tfrac{1}{2}\dot\nu(t - t_\oplus)^2 + \nu \Delta_{\text{Sh}},\end{aligned} \qquad (34)$$

where

$$\Delta_{\text{Sh}} = -\tau_M \ln\left[1 + q^2(t - T_0)^2\right]. \qquad (35)$$

The inverse function of $\Phi$ ($\Phi^{-1}$) gives the barycentric time at which the pulsar shows the phase $\phi$ :

$$t = \Phi^{-1}(\phi; \phi_0, \nu, \dot\nu, \ddot\nu, M, q, T_0). \qquad (36)$$

Having $N$ measurements $(\phi_k, t_k)$, where $\phi_k$ and $t_k$ are pulsar phase and the corresponding (barycentric) arrival

$\theta_1 \ldots \theta_n$ ($n$ = number of parameters), by the least square fit to the data:

$$\chi^2 = \sum_{k=1}^{N} \left(\frac{t_k - \Phi^{-1}(\phi_k; \theta_1 \ldots \theta_n)}{\sigma_k}\right)^2 \to \text{Min.}, \qquad (37)$$

where $\sigma_k$ is the error in the barycentric arrival time of the $k^{\text{th}}$ measurement. It is convenient to use $\Phi$ instead of $\Phi^{-1}$:

$$\chi^2 = \sum_{k=1}^{N} \left(\frac{\phi_k - \Phi(t_k; \theta_1 \ldots \theta_n)}{\nu \sigma_k}\right)^2 \to \text{Min.} \qquad (38)$$

We end this section with an investigation of the expected accuracies for derived parameters. We will introduce two length scales denoted by $\mathcal{R}_1$ and $\mathcal{R}_2$. Here $\mathcal{R}_1$ is the maximum value of the impact parameter $b_m$ that allows a reliable parameter determination and by $\mathcal{R}_2$ we denote a value of the impact parameter $b_m$ at which the travel-time delay is just visible in the residuals of a $\phi_0$-$\nu$-$\dot\nu$-fit.

For the numerical investigations we assume a 10 years observation of a 100 Hz pulsar where TOAs are taken twice a month. Moreover we assume $v_\perp = 300$ km/s, $T_0 - t_0 = 5$ yr, $\ddot\nu = 0$, and $b_m \gg R_E$ (see Eq. 3). We further assume that the pulse profile is sampled with 1024 bin and that the $1\sigma$-error for timing is 0.1 bin.

Monte Carlo simulations of synthetic data sets lead to the values given in Tables 1. For lenses below 0.5 $M_\odot$ the parameter determination is rather poor. But there is still the chance to see the effect in the residuals. We find as a good approximation

$$\mathcal{R}_2 [\text{AU}] \approx \begin{cases} 250 \mathcal{M}^{1.1} & \text{if } 0.1 \leq \mathcal{M} < 0.5, \\ 150 \mathcal{M}^{0.4} & \text{if } 0.5 \leq \mathcal{M} < 20. \end{cases} \qquad (39)$$

**Table 1.** Weak lensing event

| $\mathcal{M}$ | 0.1 | 1 | 10 | 20 |
|---|---|---|---|---|
| $b_m$ [AU] | — | 10 | 50 | 100 |
| $\Delta M/M$ | — | 10% | 5% | 7% |
| $\Delta q/q$ | — | 20% | 5% | 5% |
| $\frac{\Delta(t_\oplus - T_0)}{(t_\oplus - T_0)}$ | — | 2% | 0.5% | 0.5% |
| $\mathcal{R}_1$ [AU] | — | 30 | 100 | 200 |
| $\mathcal{R}_2$ [AU] | 20 | 150 | 350 | 500 |

This result for $\mathcal{R}_2$ was obtained for $v_\perp = 300$ km/s, though for $M \lesssim 20 M_\odot$ $\mathcal{R}_2$ proves to be not very sensitive to $v_\perp$ as long as $v_\perp \gtrsim 100$.

If the impact parameter exceeds $\mathcal{R}_2$ then the travel-time delay will not show up in the residuals. Generally speaking, this is a situation where $q$ is very small, i.e. $b_m$ is very big and/or $v_\perp$ is very small. In this case Eq. (35) reduces to

$$\Delta_{\text{Sh}} \simeq -\tau_M q^2 (t - T_0)^2. \qquad (40)$$

certain time $t$ is given by

$$\begin{aligned}
\phi &= \tilde{\phi}_0 + \tilde{\nu}(t - t_\oplus) + \tfrac{1}{2}\tilde{\dot{\nu}}(t - t_\oplus)^2, \\
\tilde{\phi}_0 &= \phi_0 - \nu \tau_M q^2 (t_\oplus - T_0)^2, \\
\tilde{\nu} &= \nu \left[ 1 - 2 \tau_M q^2 (t_\oplus - T_0) \right], \\
\tilde{\dot{\nu}} &= \dot{\nu} - 2 \nu \tau_M q^2.
\end{aligned} \qquad (41)$$

Therefore the travel-time delay is hidden in the observed period and period derivative, i.e. it is not possible to extract any information on $M$, $q$, and $T_0$ from the TOAs without making assumptions about the intrinsic frequency ($\nu$) and the intrinsic frequency change ($\dot{\nu}$) of the pulsar.

## 6. Parameter analysis for a strong lensing event ($f_m \lesssim 1$)

If the impact parameter $b$ is of the order of the Einstein radius $R_E$ we call this a *strong microlensing event*. In this case the second image is of comparable brightness, see Eq. (5), e.g. for $f = 0.5$ the second image ($-$) has approximately 40% of the intensity of the first image ($+$). Receiving two pulsar signals of the same pulsar can be used to extract more efficiently information from the observations.

The radiotelescope will receive two pulsar signals with different periods and different intensities. For the phase between the strong pulse ($+$) and the weak pulse ($-$) we find

$$\delta = \frac{\tau_- - \tau_+}{P_+} \quad \text{mod} \quad 1, \qquad (42)$$

where

$$\tau_- - \tau_+ = \tau_M \left[ f \sqrt{f^2 + 4} + 2 \ln \left( \frac{\sqrt{f^2+4}+f}{\sqrt{f^2+4}-f} \right) \right], \qquad (43)$$

see Eq. (10). (Eq. (43) was obtained in a more general form by Cooke & Kantowski 1975). Using the relation

$$\tau_- - \tau_+ = \tau_M \left( \frac{r-1}{\sqrt{r}} + \ln r \right) \qquad (44)$$

(see Krauss & Small 1991) we get $\delta$ as a function of $r$. In principle this could be used to determine the mass $M$ if $P_+ > \tau_- - \tau_+$.

### 6.1. Stellar masses

For stellar masses ($M = 1 \ldots 20 M_\odot$) $\tau_M$ is of the order 10 $\mu$s to 0.2 ms. But the minimum 10%-width of the pulse profile at 4.75 GHZ is (see Kramer et al. 1994)

$$W_{10\%} \approx 9° \times (P[\text{s}])^{-0.5}, \qquad (45)$$

which corresponds to a time width of

$$\delta t_{10\%} \approx (25 \text{ ms}) \times (P[\text{s}])^{0.5}. \qquad (46)$$

$\delta t_{10\%}$ and so one will not be able to see two separate pulses, and the measurement of $\tau_- - \tau_+$ will not be possible. There is only one pulse brightened by the factor $A$ given by Eq. (1). Only in the case of a millisecond pulsar ($\nu \gtrsim 100$ Hz) and $M \gtrsim 20 M_\odot$ we could expect a two-pulse profile.

As in the case of microlensed ordinary stars the change of $A$ with time can be used to extract some information from the observations. The problem here is that pulsar radiations shows strong luminosity variations caused by scintillation (see Ricket 1977), intrinsic instabilities, etc., which can limit the accuracy of parameter determination quite strongly.

As in Sect. 5 the most promising method to get information on the lens is the travel-time delay. But this time we have to use Eq. (10) instead of the simple Eq. (15). We get $f_m$ as a further parameter to characterize the strong microlensing event. Analogously to Sect. 5 we find

$$\Phi(t; \phi_0, \nu, \dot{\nu}, M, f_m, q, T_0) = \phi_0 + \nu(t - t_\oplus) + \tfrac{1}{2} \dot{\nu}(t-t_\oplus)^2 + \nu \Delta_+, \qquad (47)$$

where

$$\begin{aligned}
\Delta_+ &= \tau_M \left( 4 F_+^{-2} - 2 \ln F_+ \right), \\
F_+ &= \sqrt{f^2 + 4} + f, \quad f = f_m \sqrt{1 + q^2(t-T_0)^2}.
\end{aligned} \qquad (48)$$

Similarly to Sect. 5 we performed Monte Carlo simulations of synthetic data sets to estimate the expected accuracies in the parameter determination. We arrived at the conclusion that for stellar masses it will not be possible to get a reliable fit for all seven parameters ($\phi_0, \nu, \dot{\nu}, M, f_m, q, T_0$) unless timing accuracy is increased by at least an order of magnitude. Especially the parameters $f_m$ and $q$ appear to have rather large errors. The most practicable method for $M \sim 10 M_\odot$ and $f_m \lesssim 0.5$ is to limit $f_m$ by the determination of the maximum flux magnification $A_{\max} = A(f_m)$, see Eqs. (1), and then fit for the other parameters. Table 2 shows the result for a 5 years weekly observation of a 100 Hz pulsar around the maximum of the microlensing event ($\dot{\nu} = -10^{-15}\text{s}^{-2}$, $v_\perp = 300$ km/s). We assumed an accuracy of 20% for the measurement of the maximum flux magnification $A_{\max}$.

**Table 2.** Strong microlensing event: $M = 10 M_\odot$, $R_E = 10$ AU

|  | $b_m = 2$ AU | $b_m = 5$ AU |
|---|---|---|
| $f_m$ | $0.17 \ldots 0.25$ | $0.40 \ldots 0.66$ |
| $\Delta M / M$ | 1% | 2% |
| $\Delta q / q$ | 16% | 13% |
| $\frac{\Delta(t_\oplus - T_0)}{(t_\oplus - T_0)}$ | 0.2% | 0.2% |

In case of a sun-like mass the strong travel-time delay will clearly show up in the residuals. But even if we use $A_{\max}$ to extract $f_m$ from the observations, there will be no we can restrict the error in $f_m$ to about 20%, Monte Carlo simulations show that there will be nearly no restrictions on the parameter $q$. If we restrict $q$ by some plausible a priori assumptions, e.g. about $v_\perp$, we can still extract acceptable values for $M$ and $T_0$ from the data with an error that does depend only very little on the error made in the a priori assumption.

### 6.2. A SMBH in the Galactic Center

For a SMBH of $10^6 M_\odot$ hosted in the very center of the Galaxy we find the Einstein radius to be several orders bigger than for stellar masses

$$R_E \approx (2900 \text{ AU}) \left(\frac{D_{Mp}}{1 \text{ kpc}}\right)^{1/2}. \qquad (49)$$

But a pulsar located a few 100 pc behind the GC and taking part in the Galactic rotation will have a transverse velocity of $v_\perp \sim 150$ km/s (see Kent 1992). Therefore the pulsar will cross in the lens plane a path of 32 AU in one year, which is by far smaller than the Einstein radius. Therefore $f$ will change only by a small amount $\Delta f$ during an observation of a few years and we need $f \lesssim 1$ to see a noticeable change during such an observation. Thus we can expect to see both images where $1 \leq r \lesssim 10$. Recording a typical "bell-shaped" curve (see Fig. 3) in the timing data would require an observation of about a 100 years. In Table 3 we present the changes of various flux and time parameters during the last 2 years before reaching the minimum of the impact parameter ($t = T_0 - 2 \text{ yr} \ldots T_0$), for a pulsar located a 100 pc behind the SMBH.

**Table 3.** SMBH: $M = 10^6 M_\odot$, $R_E = 1000$ AU

| $b_m$ [AU] | $\mu_+$ | $\mu_-$ | $r$ |
|---|---|---|---|
| 500 | 1.58…1.59 | 0.58…0.59 | 2.71…2.69 |
| 200 | 2.93…3.04 | 1.93…2.04 | 1.52…1.49 |
| 50 | 6.72…10.5 | 5.72…9.52 | 1.17…1.11 |

| $b_m$ [AU] | $\Delta\tau_+$ | $\Delta\tau_-$ | $\tau_- - \tau_+$ |
|---|---|---|---|
| 500 | 61 ms | -101 ms | 20.04…19.88 s |
| 200 | 173 ms | -213 ms | 8.27…7.88 s |
| 50 | 583 ms | -623 ms | 3.18…1.97 s |

The following three consequences can be extracted from Table 3:

- There is practically no variation in the ratio of the intensities of the two images within years of observation for all values of $b_m$. We can safely use $r \simeq const$.
- Only for small values of $f$ there is a measurable change in $\mu_+$ and $\mu_-$.
- $\tau_- - \tau_+$ is in general much greater than a typical period of a pulsar. Therefore it is not possible to read off $\tau_- - \tau_+$ from the pulse profile. Consequently we cannot some kind of distinct outbursts in the pulse emission.

The first two consequences imply that there is only very little chance to extract information from the measured intensities of the pulsar signals. For $b_m \lesssim 1000$ AU we can use Eqs. (5) to extract a crude estimation for $f$ from the observed intensities. The corresponding error is

$$\Delta f = \frac{\sqrt{f^2 + 4}}{4} \frac{\Delta r}{r}. \qquad (50)$$

Much more promising is the use of Eq. (43). Here we have to measure the phase $\delta$ as a function of time. Figure 6 illustrates the change of the observed pulse profile within a period of two years before $T_0$.

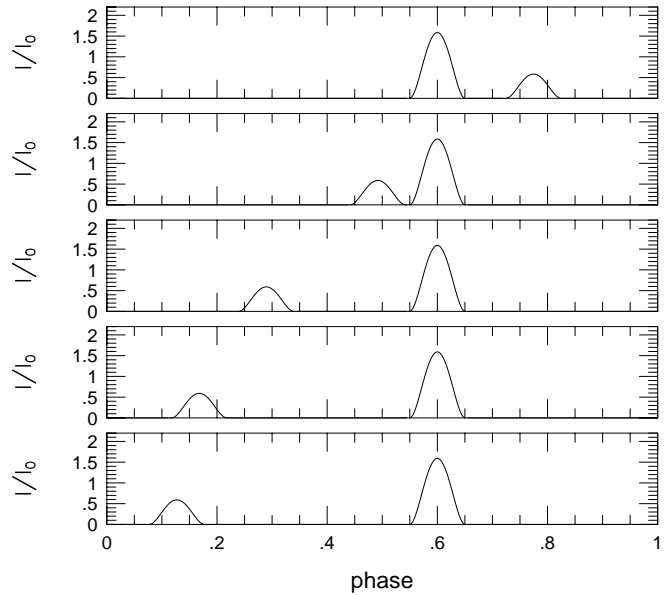

**Fig. 6.** Change of the pulse profile due to gravitational lensing by a SMBH ($M = 10^6 M_\odot$) in the GC. The pulsar has a period of 0.25 s and is located 100 pc behind the GC. $b_m = 500$ AU, $v_\perp = 150$ km/s. The pulse profiles are shown in steps of 0.5 years where the last pulse profile corresponds to the time $T_0$.

Monte Carlo simulations of synthetic data sets show that by fitting

$$P_+ \delta(t) = \tau_+ - \tau_- - N_\delta P_+, \qquad N_\delta \in \mathbb{N} \qquad (51)$$

to the observations we have to distinguish three cases:

I) Eq. (43) can be approximated by

$$P_+ \delta(t) = 2\tau_M f_m q^2 (t - T_0)^2 + (4\tau_M f_m - N_\delta P_+). \quad (52)$$

In general, $b_m$ has to be a few times larger than $R_E$. But then $\mu_- \ll 1$ and therefore in practice we will not be able to see the second image. Eq. (52) does not allow mass determination.

to fit all five parameters $(M, f_m, q, T_0, N_\delta)$. Here we use Eq. (5) to get a crude estimation for the value of $f_m = f(T_0)$. (The time $T_0$ can be extracted quite easily from the observation data with sufficient high accuracy.) We will use this result in Eq. (51) and then fit for the parameters $M, q, T_0, N_\delta$. An error of 25% in $f_m$ leads typically to an error of a 10...20% in the determination of the mass $M$.

III) $b \lesssim 500$ AU. We have to use the full Eq. (43) to fit the observed data. Table 4 shows the expected accuracy for the mass determination for a 10-years monthly observation of a 250 ms pulsar, $t_0 - T_0 = -5$ yr.

**Table 4.** SMBH: $M = 10^6 M_\odot$, $R_E = 1000$ AU

| $b_m$ [AU] | 500 | 250 | 100 | 50 | 10 |
|---|---|---|---|---|---|
| $\Delta M/M$ | 10% | 4% | 3% | 2% | 1.5% |

Concerning cases II and III it is important to note that on one hand the period of the pulsar is very well known, and on the other hand the quantity $N_\delta$ is an integer number. In case of long-period pulsar $N_\delta$ can be determined with rather high confidence, and thus it can be advantageous to use Eqs. (51) and (44) to determine the mass, especially in case II.

As in the case of stellar masses we can also use the TOAs of the brighter pulse (+) to do a parameter analysis by fitting Eq. (47). Monte Carlo simulations of synthetic data sets show that this method is practicable for $b_m \lesssim 300 AU$. The expected accuracies are worse by more than a factor of two compared with the accuracies achieved in the previous method, i.e. by fitting $\delta(t)$.

The influence of the travel-time delay can be seen in the residuals of a 10 years observation of a 10 Hz pulsar up to an impact parameter $b_m$ of about 2000 AU (see Fig. 7), and thus $\mathcal{R}_2 \approx 2000$ AU for a $10^6 M_\odot$ SMBH in the Galactic center. Only for $b_m \gtrsim 2000$ AU fitting Eq. (47) with $\tau_M \equiv 0$ would provide a good fit.

## 7. super-massive black hole versus super dense star cluster

So far we have payed attention only to the Schwarzschild masses as candidates for lenses. Since at the present time it is not clear whether the very center of our Galaxy hosts a SMBH or a super-dense star cluster (SDSC) we will investigate the possibility of distinguishing between these two models by observing a lensed pulsar in direction of the Galactic nucleus.

As mentioned in the Introduction, observations indicate that a mass of $10^6 M_\odot$ is resident within a 0.1 pc ($\sim 20\,000$ AU) radius around the very center of the Galaxy. This corresponds to a mass density of about

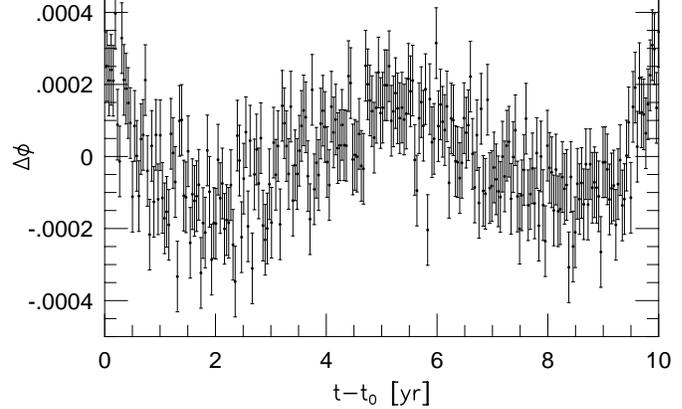

**Fig. 7.** Residuals of a $\phi_0 - \nu - \dot\nu$-fit for a 10 years observation of a pulsar that is lensed by a $10^6 M_\odot$ SMBH in the GC where $b_m = 2000$ AU, $v_\perp = 150$ km/s, $R_E = 1000$ AU, and $T_0 - t_0 = 5$ yr.

$10^9 M_\odot/\text{pc}^3$. We use Plummer's law (see Binney & Tremaine 1987)

$$\varrho(R) = \frac{\varrho_0}{[1 + (R/R_*)^2]^{5/2}} \tag{53}$$

to model the density distribution of the putative SDSC. Here $\varrho_0$ is the central mass density and $R_*$ is a length that characterizes the size of the SDSC. Let us denote the total mass of the star cluster by $M_*$. Integration of Eq. (53) leads to

$$M_* = \tfrac{4}{3}\pi \rho_0 R_*^3. \tag{54}$$

About 85% of $M_*$ are within a radius of $3R_*$. The corresponding surface mass density is

$$\Sigma(\xi) = \frac{\Sigma_0}{[1 + (\xi/R_*)^2]^2}, \tag{55}$$

where

$$\Sigma_0 = \tfrac{4}{3}\rho_0 R_* = \frac{1}{\pi}\frac{M_*}{R_*^2} \tag{56}$$

is the central surface mass density. If $\Sigma_0$ is smaller than the critical surface mass density

$$\Sigma_{cr} = \frac{c^2}{4\pi G}\frac{D_p}{D_M D_{Mp}}, \tag{57}$$

we will observe only one image of the pulsar. If $\Sigma_0 > \Sigma_{cr}$ we will have one or three images, depending on $b$ (see Schneider et al. 1991). For a lensing event in the GC ($D_p \approx D_M$) we find

$$\Sigma_0/\Sigma_{\rm cr} \approx 1.9 \times 10^{-7}\left(\frac{M_*}{10^6 M_\odot}\right)\left(\frac{R_*}{1\,\text{pc}}\right)^{-2}\left(\frac{D_{Mp}}{1\,\text{pc}}\right). \tag{58}$$

well below 1 for $D_{Mp} \lesssim 1$ kpc and thus we expect only one image for all impact parameters $b$.

To shorten notation and facilitate the comparison between the SMBH and the SDSC we define

$$\tilde{R}_S \equiv \frac{2GM_*}{c^2}, \qquad \tilde{R}_E \equiv \sqrt{2\tilde{R}_S \frac{D_M D_{Mp}}{D_p}}. \qquad (59)$$

We define the dimensionless parameter

$$\kappa_0 \equiv \Sigma_0/\Sigma_{\rm cr} = \tilde{R}_E^2/R_*^2. \qquad (60)$$

For small impact parameters ($b \ll R_*$) the lens equation for Plummer's model leads to (see Appendix A)

$$\xi = \frac{b}{1-\kappa_0}. \qquad (61)$$

The corresponding travel-time delay is (see Appendix A)

$$\tau = -\tau_{M_*}\left(\kappa_0^{-1} - 1\right) \tilde{f}^2 + const., \qquad (62)$$

where we used the definitions

$$\tau_{M_*} \equiv \tilde{R}_S/c, \qquad \tilde{f} \equiv b/\tilde{R}_E. \qquad (63)$$

Since

$$\tilde{f} = \tilde{f}_m \sqrt{1+s^2}, \qquad s = q(t-T_0) \qquad (64)$$

we obtain

$$\tau = -\mathcal{C}_*(t-T_0)^2 + const.,$$
$$\mathcal{C}_* \equiv \frac{\tau_{M_*}}{1-\kappa_0}\left(\frac{v_\perp}{R_E}\right)^2. \qquad (65)$$

Thus, in contrast to the SMBH,

$$\phi = \tilde{\phi}_0 + \tilde{\nu}(t-t_\oplus) + \tfrac{1}{2}\tilde{\dot{\nu}}(t-t_\oplus)^2 \qquad (66)$$

would fit the observed TOAs. We find

$$\begin{aligned}
\phi &= \tilde{\phi}_0 + \tilde{\nu}(t-t_\oplus) + \tfrac{1}{2}\tilde{\dot{\nu}}(t-t_\oplus)^2, \\
\tilde{\phi}_0 &= \phi_0 - \nu \mathcal{C}_*(t_\oplus - T_0)^2, \\
\tilde{\nu} &= \nu\left[1 - 2\mathcal{C}_*(t_\oplus - T_0)\right], \\
\tilde{\dot{\nu}} &= \dot{\nu} - 2\nu \mathcal{C}_*.
\end{aligned} \qquad (67)$$

For a SDSC with $M_* = 10^6 M_\odot$, $R_* = 0.1$ pc, $\tilde{R}_E = 1000$ AU, and $v_\perp = 150$ km/s one finds

$$\dot{\nu} - \tilde{\dot{\nu}} = \nu(2.0 \times 10^{-17} {\rm s}^{-1}). \qquad (68)$$

Since $\dot{\nu} \sim -10^{-15}$ for most pulsars, one does not expect to discover the influence of a SDSC in the TOAs. Note: The influence of the SDSC is independent from $b_m$.

Based on the numerical investigations in Sect. 5 and Sect. 6 we give some coarse estimations for the probability of microlensing of pulsar radiation in the GC. (See also Krauss & Small 1991).

### 8.1. Luminous matter

To estimate the lensing probability by ordinary stars we shall calculate the expected number $N$ of stars between the pulsar and the Earth which are able to influence noticeably the TOAs.

For the density of luminous sources in the Galactic plane we use an empirical model of Bahcall & Soniera (1980):

$$\begin{aligned}
\varrho(r) &= \varrho_{\rm disk}(r) + \varrho_{\rm bulge}(r) \\
&= \left(4 \times 10^7 \frac{M_\odot}{{\rm kpc}^3}\right) \exp\left[\frac{R_0 - r}{3.5\,{\rm kpc}}\right] \\
&\quad + \left(3 \times 10^9 \frac{M_\odot}{{\rm kpc}^3}\right) \left(\frac{r}{1\,{\rm kpc}}\right)^{-1.8} \exp\left[-\left(\frac{r}{1\,{\rm kpc}}\right)^3\right].
\end{aligned} \qquad (69)$$

Here $\varrho_{\rm disk}$ and $\varrho_{\rm bulge}$ are the contributions of the disk and the bulge, respectively, $R_0$ is the distance of the Sun from the Galactic Center ($\approx 8$ kpc) and $r$ is the distance from the Galactic Center. $\varrho_{\rm bulge}(r)$ is not valid for the very central region ($r \lesssim 1$ pc).

The number density of the stars that have a mass between $\mathcal{M}$ and $\mathcal{M} + d\mathcal{M}$ can be estimated using the initial mass function given in Kroupa (1995). We find

$$\xi(\mathcal{M}, r) d\mathcal{M} \approx$$
$$1.4 \left(\frac{\varrho(r)}{M_\odot}\right) d\mathcal{M} \begin{cases} 0.5^{1.3} \mathcal{M}^{1.3}, & {\rm if}\ 0.08 \leq \mathcal{M} < 0.5, \\ 0.5^{2.2} \mathcal{M}^{-2.2}, & {\rm if}\ 0.5 \leq \mathcal{M} < 1.0, \\ 0.5^{2.2} \mathcal{M}^{-2.7}, & {\rm if}\ 1.0 \leq \mathcal{M} < \infty. \end{cases} \qquad (70)$$

Thus the number of stars that influence noticeably the TOAs is

$$N = \int_0^{D_Q} \int_{0.08}^\infty \pi[\mathcal{R}_2(\mathcal{M})]^2 \xi(\mathcal{M}, l)\, d\mathcal{M}\, dl, \qquad (71)$$

where $l$ parameterizes the pulsar–Earth distance. The density $\varrho$ is given by Eq. (69) and the radius $\mathcal{R}_2$ we take from Sect. 5, Eq. (39). Integration in Eq. (71) over the mass $\mathcal{M}$ leads to

$$N = 1.6 \times 10^{-12} \int_0^{D_Q[{\rm kpc}]} \varrho(l)[M_\odot/{\rm kpc}^{-3}]\, dl. \qquad (72)$$

Figure 8 shows the resulting number $N$ for a pulsar located at $r = 0.5$ kpc as function of the angle between the directions GC–SSB and GC–PSR. Strictly speaking, Fig. 8 is valid for the observation of a 100 Hz pulsar where $v_\perp \sim 300$ km/s was assumed for all the lenses. For the majority of the pulsars, which has frequencies well below 100 Hz, the number $N$ is smaller by an order of magnitude.

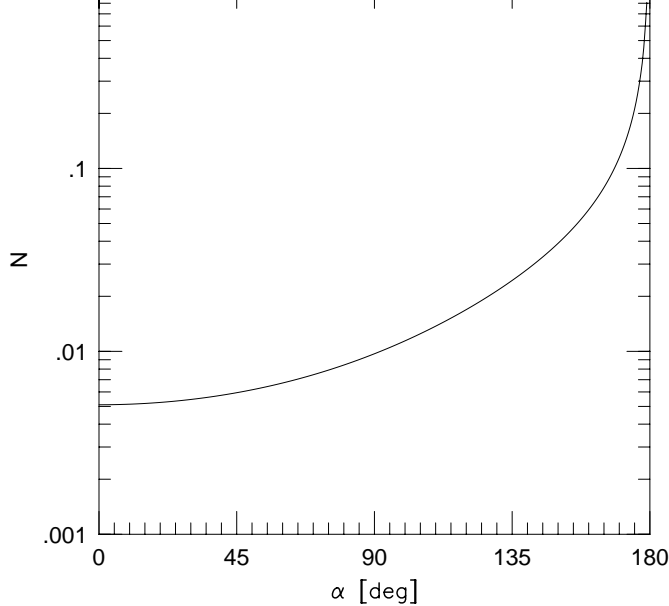

**Fig. 8.** Expected number of lenses $N$ for a 100 Hz pulsar at 0.5 kpc distance from the GC as function of the angle $\alpha$ between the directions GC–SSB and GC–PSR. $v_\perp \sim 300$ km/s is assumed.

Figure 8 suggests that for pulsars located behind the GC there is a high probability that the timing data are influenced by some travel-time delay. If a considerable part of the dark matter in our Galaxy exists in form of massive objects ($M \gtrsim 0.5 M_\odot$) the lensing probability is increased by an order of magnitude.

### 8.2. The SMBH model

Finally we would like to estimate the expected number of pulsars behind a putative SMBH in the GC. For a SMBH of $10^6 M_\odot$ we have $\mathcal{R}_1 \sim 1000$ AU ($\mathcal{R}_2 \sim 2000$ AU). Using Eq. (69) we find a mass of 70 $M_\odot$ for $\mathcal{R}_1$ (280 $M_\odot$ for $\mathcal{R}_2$) between 1 pc and 1 kpc behind the SMBH. If we assume that 0.1% of the mass is provided by neutron stars, then there should be approximately 0.07 (0.3) neutron stars behind the SMBH. If 10% of these neutron stars are visible as pulsars, we can expect more than 0.007 (0.03) pulsars suffering an observable Shapiro delay caused by the SMBH. This number is further decreased by the fact that one expects to find mostly the young pulsars in the GC, which are both bright and have wider beams.

### 9. Conclusions

The discovery of a large number of pulsars in the GC seems to be only a matter of time. The GC region contains pro- and, perhaps, even a SMBH. In this paper we consider the relativistic interaction of pulsar radiation with a moving massive object. Based on extensive Monte Carlo simulations of synthetic data sets we argue, that the important new effect to be taken into account in pulsar timing is the travel-time delay. This could lead to a very good determination of the mass of the lens as it was shown in Sect. 5 and Sect. 6. The statistical estimates of Sect. 8 imply that the probability for observing a travel-time event with a lens mass between 1 and 20 $M_\odot$ is quite reasonable due to the high density of stars in the galactic bulge and the large radius of influence $\mathcal{R}_2$ even for small masses.

In Sect. 6 it was shown that the travel-time delay could provide a unique possibility for the detection of a SMBH in the GC. Regular timing observations over a period of a few years would lead to a mass determination with an accuracy of a few percent. As calculated in Sect. 7 the travel-time delay caused by a SDSC in the Galactic nucleus will lead to TOAs which are indistinguishable from the TOAs of an unlensed isolated pulsar and thus they are quite different from those of a SMBH. Thus the detection of a lensed pulsar in the very center of the Galaxy could bring to the end the SMBH-SDSC-controversy. Unfortunately the situation having a pulsar within a radius of a 1000 AU around the galactic nucleus (Sag A$^*$) seems to be not very likely.

*Acknowledgements.* We are grateful to Gerhard Schäfer for useful comments and careful reading of the manuscript. It is our pleasure to thank Oleg Doroshenko, Wilhelm Kley, Michael Kramer, and Andrew Lyne for valuable discussions. JG and MS are grateful to the Alexander for Humboldt Foundation for donating the SPARC 10/40 work station and the Max-Planck Institute for Radioastronomy for the loan of the computer system. This paper was partially supported by the Grant of Polish Committee for Scientific Research KBN PB 2P30400305.

### A. The travel-time delay

By *travel-time delay* of a light ray we denote the (coordinate) time delay of this light ray relative to the undeflected light ray propagating in an Euclidean background metric. The travel-time delay for the light ray that intersects the lens plane at the position $\boldsymbol{\xi}$ is given by (see e.g. Schneider et al. 1991)

$$c\tau = \frac{D_M D_{Mp}}{2 D_p}[\hat{\boldsymbol{\alpha}}(\boldsymbol{\xi})]^2 - \hat{\psi}(\boldsymbol{\xi}) + const. \quad , \tag{A1}$$

where the deflection potential $\hat{\psi}$ and the deflection angle $\hat{\boldsymbol{\alpha}}$ are given by

$$\hat{\psi}(\boldsymbol{\xi}) = \frac{4G}{c^2} \int d^2\xi' \, \Sigma(\boldsymbol{\xi}') \, \ln\left(\frac{|\boldsymbol{\xi} - \boldsymbol{\xi}'|}{\xi_0}\right) \tag{A2}$$

and

$$\hat{\boldsymbol{\alpha}}(\boldsymbol{\xi}) = \nabla \hat{\psi}(\boldsymbol{\xi}), \tag{A3}$$

density of the deflector onto the lens plane and $\xi_0$ is a typical length scale in the lens plane.

For axially symmetric lenses, i.e. $\Sigma(\boldsymbol{\xi}) = \Sigma(|\boldsymbol{\xi}|)$, we find (see e.g. Schneider et al. 1991)

$$\hat{\psi}(\xi) = \frac{8\pi G}{c^2} \int_0^\xi d\xi'\, \xi'\, \Sigma(\xi')\, \ln\left(\frac{\xi}{\xi'}\right) + const. \qquad (A4)$$

and

$$\hat{\alpha}(\xi) = \frac{8\pi G}{c^2 \xi} \int_0^\xi d\xi'\, \xi'\, \Sigma(\xi'). \qquad (A5)$$

### A.1. The Schwarzschild lens

If the deflected light ray propagates in a Schwarzschild spacetime characterized by the mass $M$ then

$$\Sigma(\boldsymbol{\xi}) = M\delta^2(\boldsymbol{\xi}). \qquad (A6)$$

We obtain

$$\hat{\psi}(\xi) = \frac{4GM}{c^2} \ln\left(\frac{\xi}{R_E}\right) \qquad (A7)$$

and

$$\hat{\alpha}(\xi) = \frac{4GM}{c^2 \xi}. \qquad (A8)$$

Since

$$\xi = \xi_\pm = \frac{R_E}{2} F_\pm, \qquad F_\pm \equiv \sqrt{f^2 + 4} \pm f \qquad (A9)$$

we obtain further

$$c\tau_\pm = R_S \left(4 F_\pm^{-2} - 2\ln F_\pm\right) + const. \qquad (A10)$$

### A.2. Plummer's model

If our deflecting mass is a spherically symmetric star cluster with a density distribution that follows Plummer's law then the corresponding surface mass density is given by

$$\Sigma(\xi) = \frac{\Sigma_0}{[1 + (\xi/R_*)^2]^2}, \qquad \Sigma_0 = \frac{1}{\pi}\frac{M_*}{R_*^2}. \qquad (A11)$$

Using Eqs. (A4),(A5) we obtain

$$\hat{\psi}(\xi) = \tilde{R}_S \ln[1 + (\xi/R_*)^2] + const. \qquad (A12)$$

and

$$\hat{\alpha}(\xi) = 2\tilde{R}_S \frac{\xi}{R_*^2 + \xi^2}. \qquad (A13)$$

The lens equation is

$$b = \xi - \frac{D_M D_{Mp}}{D_p} \hat{\alpha}(\xi) = \xi \left(1 - \frac{\tilde{R}_E^2}{R_*^2 + \xi^2}\right). \qquad (A14)$$

i.e. $\tilde{R}_E < R_*$, then Eq. (A14) has only one solution. For small impact parameters $b$ $(\ll R_*)$ we find

$$\xi = \frac{b}{1 - \kappa_0}, \qquad \kappa_0 = \frac{\tilde{R}_E^2}{R_*}. \qquad (A15)$$

Consequently, using Eq. (A1), we find

$$c\tau = -\tilde{R}_S (1 - \kappa_0)^{-1} \tilde{f}^2 + const., \qquad \tilde{f} \equiv b/\tilde{R}_E. \qquad (A16)$$